\documentclass[prb,twocolumn,showpacs,byrevtex,altaffillsymbol,superscriptaddress,citeautoscript,amsfonts,amssymb,amsmath]{revtex4}

\usepackage{graphicx}
\usepackage{dcolumn}
\usepackage{amsmath}

\begin{document}

\date{\today}

\title{Investigation of the ferromagnetic transition in the correlated 4$d$ perovskites SrRu$_{1-x}$Rh$_x$O$_3$}

\author{K. Yamaura}
\email[E-mail at:]{YAMAURA.Kazunari@nims.go.jp}
\homepage[Fax.:]{+81-29-860-4674}
\affiliation{Superconducting Materials Center, National Institute for Materials Science, 1-1 Namiki, Tsukuba, Ibaraki 305-0044, Japan}

\author{D.P. Young}
\affiliation{Department of Physics and Astronomy, Louisiana State University, Baton Rouge, LA 70803}

\author{E. Takayama-Muromachi}
\affiliation{Superconducting Materials Center, National Institute for Materials Science, 1-1 Namiki, Tsukuba, Ibaraki 305-0044, Japan}

\begin{abstract}
The solid-solution SrRu$_{1-x}$Rh$_x$O$_3$ ($0\le x \le1$) is a variable-electron-configuration system forming in the nearly-cubic-perovskite basis, ranging from the ferromagnetic 4$d^4$ to the enhanced paramagnetic 4$d^5$. 
Polycrystalline single-phase samples were obtained over the whole composition range by a high-pressure-heating technique, followed by measurements of magnetic susceptibility, magnetization, specific heat, thermopower, and electrical resistivity. 
The ferromagnetic order in long range is gradually suppressed by the Rh substitution and vanishes at $x \sim 0.6$. 
The electronic term of specific-heat shows unusual behavior near the critical Rh concentration; the feature does not match even qualitatively with what was reported for the related perovskites (Sr,Ca)RuO$_3$.
Furthermore, another anomaly in the specific heat was observed at $x \sim 0.9$.

\end{abstract}


\maketitle

\section{Introduction}
Perovskite-ruthenium oxides and the single-structure family form a distinguished class of the 4$d$-band oxides that show rich electronic phenomena, such as unconventional superconductivity \cite{PRB97KI,PRL01KI}, a quantum critical transition \cite{SCIENCE01SAG,PRL01RSP,PRB01DJS,PRB00SII,PRL98AVP}, and negative spin polarization \cite{PRL00DCW}. 
These phenomena are of great interest to the condensed matter community, and thus, these materials have recently received much attention.
In regards to the nearly cubic perovskite SrRuO$_3$, many efforts have focused on exploring its itinerant ferromagnetism.
The electrically conducting state of SrRuO$_3$ is probably dominated by Ru-$4d$ electrons and a small amount of O-$2p$ holes \cite{PRL01MSL,JAP96DJS}, and is clearly affected by the ferromagnetic order ($T_{\rm c}\sim$160 K) \cite{JPSJ78AK,JPSJ76AK,JPSJ76AK2,JAP68JML,IC66AC}. 
So far experimental studies on the material reveal strongly correlated features and substantial influences of spin fluctuations through measurements of optical conductivity \cite{PRB01JSL}, far-infrared transmission \cite{PRL00JSD}, Ru-NMR \cite{PRB99HM}, photoemission spectra \cite{PRB99JO}, magneto-optical conductivity \cite{PRB99JSD}, reflectivity \cite{PRL98PK}, magnetization and resistivity of thin films \cite{PRL00RR,PRL00LK,PRL96LK,JPCM96LK}, and measurements of the Hall coefficient \cite{PRB96PBA}. 
Those data clearly indicate that the perovskite SrRuO$_3$ goes beyond our understanding of normal metals. 

In a sharp contrast with previous reports, a highly pure single crystal of SrRuO$_3$, which was less influenced by disorder, was found to show normal characteristics within Fermi-liquid theory \cite{PRL02LC}.
This fact probably indicates that the electrically conducting state is highly sensitive to disorder as mentioned in the report. 
Regardless of the influence of disorder, we were motivated to chemically dope ferromagnetic SrRuO$_3$ over a wide range to explore a dramatic transition to a novel electronic state, as was often achieved in this manner. 

Recently, the synthesis of SrRhO$_3$ was reported for the first time \cite{PRB01KY}. 
The Rh oxide perovskite is structurally analogous to SrRuO$_3$, and is electrically conducting as well as SrRuO$_3$ \cite{PRB01KY}. 
The perovskite SrRhO$_3$ shows an enhanced paramagnetism \cite{PRB01KY,PB03KY}. 
Each Rh$^{4+}$ (4$d^5$) is coordinated by six oxygen atoms and is in the low-spin state due to the relatively large 10$Dq$ \cite{PRB03YSL}. 
Because the state ($t_{\rm 2g}^5e_{\rm g}^0$) forms  an $S=$ 1/2 system, appearance of possible quantum characteristics could be expected for SrRhO$_3$ and closely related materials \cite{PRB03DJS}. 
We then pursued a systematic substitution of Rh for Ru of the perovskite. 
As a result, we have succeeded in obtaining polycrystalline samples over the full solid-solution range between SrRuO$_3$ and SrRhO$_3$, which establishes a new avenue for probing the itinerant ferromagnetic state (4$d^4$) from the paramagnetic state (4$d^5$). 

In this paper, we report the magnetic and the electrical transport properties of the nearly cubic perovskites SrRu$_{1-x}$Rh$_x$O$_3$ ($0\le x \le1$), and compare the data to those of the related perovskites (Sr,La)RuO$_3$ \cite{JPCM02HN,JSSC72RJB} and (Sr,Ca)RuO$_3$ \cite{PRB00TH,PRL99KY,JPSJ98TK,PRB97GC,JAP68JML,IC66AC}. 
As expected with increasing Rh concentration in Sr(Ru,Rh)O$_3$, the ferromagnetic order is gradually suppressed and then finally disappears. 
The data clearly indicate that the electronic term of the specific heat is correlated with the magnetic character in an unusual fashion. 

\section{Experimental}
Polycrystalline single-phase samples were prepared as follows. 
Fine and pure ($>$99.9\%) powder of SrO$_2$ \cite{SrO2}, RuO$_2$, RhO$_2$, Rh, and Ru were employed as starting materials, and appropriate amounts were mixed into the stoichiometry SrRu$_{1-x}$Rh$_x$O$_3$ (0$\le x \le$1 at 0.1 step). 
Each 0.4-g sample was placed into a platinum capsule and then heated at 1500--1600 $^\circ$C for 1 hr at 6 GPa in a high-pressure furnace. 
The elevated temperature was 1550 $^\circ$C for the $x=$ 0.0 and 0.1 samples, 1500 $^\circ$C for the $x=$ 0.2 to 0.8 samples, and 1600 $^\circ$C for the $x=$ 0.9 and 1.0 samples. 
The applied pressure was held constant during the heating \cite{PRB01KY,PRB02KY} and maintained until after the capsule was quenched to room temperature. 
The polycrystalline samples were black in color and retained a pellet shape (approximately 6 mm in diameter and 2 mm in thickness). 
Each face of the sample pellet was polished carefully in order to remove any possible contamination from chemical reactions with the capsule. 

\begin{figure}
\includegraphics[width=8cm]{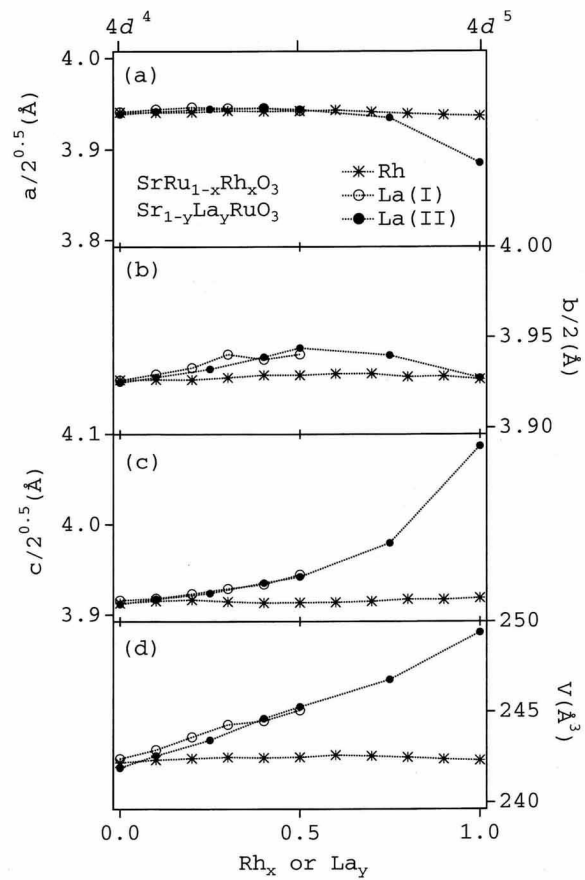}
\caption{Orthorhombic unit-cell parameters (a--c) and the unit-cell volume (d) of the perovskites SrRu$_{1-x}$Rh$_x$O$_3$ (stars) and Sr$_{1-y}$La$_y$RuO$_3$ (open and filled circles). 
The sets of data for Sr$_{1-y}$La$_y$RuO$_3$ were taken from published reports \cite{JPCM02HN,JSSC72RJB}.} 
\label{fig1}
\end{figure}

A small piece of each product was investigated by standard x-ray (CuK$\alpha$) diffraction at room temperature. 
Each x-ray pattern showed clearly a GdFeO$_3$-type-peak distribution and absence of significant impurities as well as that for the previous sample of SrRhO$_3$ \cite{PRB01KY}. 
The set of x-ray patterns indicated the formation of a full solid-solution range between SrRuO$_3$ ($Pbnm$, $a=$ 5.5670(1) \AA, $b=$ 5.5304(1) \AA, $c=$ 7.8446(2) \AA, and $V=$ 241.5 \AA$^3$~\cite{ACSC89CWJ}) and SrRhO$_3$ ($Pnma$, $a=$ 5.5394(2) \AA, $b=$ 7.8539(3) \AA, $c=$ 5.5666(2) \AA, and $V=$ 242.2 \AA$^3$~\cite{PRB01KY}). 
The orthorhombic-unit-cell volume was calculated from the unit-cell parameters deduced from the x-ray data, and those are plotted versus composition in Fig.\ref{fig1}.
Anomalies do not appear in the lattice-parameter plots, and the unit-cell volume varies more or less monotonically, indicating that the perovskite-solid solution is likely to preserve the orthorhombic symmetry over the full solid-solution range studied. 
The relatively small degree of variation of the parameters probably reflects mainly the small difference in ionic radii [$^{\rm VI}$Rh$^{4+}$ ($\sim$0.60\AA) is comparable in size to $^{\rm VI}$Ru$^{4+}$ ($\sim$0.62\AA)] \cite{ACSA76RDS}. 

Electronic properties of the samples were investigated by various ways; magnetic susceptibility, magnetization, electrical resistivity, specific heat, and thermopower were studied. 
Each data set was collected on the same sample batch. 
The magnetic properties were studied in a commercial apparatus (Quantum Design, MPMS-XL system) between 2 K and 390 K below 70 kOe. 
The electrical resistivity was measured in a commercial apparatus (Quantum Design, PPMS system) by a conventional ac-four-terminal technique; the ac-gauge current at 30 Hz was 1 mA for the $x=0$ to 0.2 samples and 0.1 mA for the $x=0.3$ to 1 samples. 
Prior to the electrical measurements, each sample pellet was cut out into a bar shape, and each face was polished with aluminum-oxide-lapping film in order to reduce the contact resistance, followed by deposition of gold pads ($\sim$200 nm in thickness) at four locations along the bar.
Silver epoxy was used to fix fine platinum wires ($\sim$ 30 $\mu$m$\phi$) at the each gold terminal. 
The contact resistance was less than 4 ohms. 
Specific-heat measurements were conducted on a small piece of each pellet (10--25 mg) in the PPMS system with a time-relaxation method over the temperature range between 1.8 K and 10.3 K. 
The $x=0.9$ sample was measured up to 20 K in the magnetic field of 70 kOe. 
The magnetic-field contribution to the background was found being smaller than 1\% of the specific heat data, and then carefully subtracted. 
Thermopower for all the samples was measured in the PPMS system between 5 K and 300 K with a comparative technique using a constantan standard. 

\section{Results and Discussion} 
Lattice parameters and the volume of the orthorhombic unit cell of SrRu$_{1-x}$Rh$_x$O$_3$ are compared with those of the isoelectronic and isostructural (Sr,La)RuO$_3$ in Fig.\ref{fig1}. 
The La data were quoted from two independent reports \cite{JPCM02HN,JSSC72RJB}.
As shown in the plots, the two sets of La data are entirely consistent with each other, and clearly reveal that the La substitution for Sr remarkably changes the unit-cell size.
In the La plots an anisotropic change of the unit-cell appears (approximately -1.4\% along the $a$-axis, nearly zero along the $b$-axis, and 4.5\% along the $c$-axis), which might reflect a change in the Ru--O-bond distances as well as the degree of cooperative rotations of the RuO$_6$ octahedra \cite{JPCM02HN,JSSC72RJB}. 
These features should result from a balance between two opposing factors in the GdFeO$_3$-type basis: the ionic-size issue [$^{\rm XII}$La$^{3+}$ ($\sim$1.36\AA) is smaller than $^{\rm XII}$Sr$^{2+}$ ($\sim$1.44\AA)] and the charge neutrality [$^{\rm VI}$Ru$^{3+}$ ($\sim$0.68\AA) is larger than $^{\rm VI}$Ru$^{4+}$ ($\sim$0.62\AA)] \cite{ACSA76RDS}. 

The unit-cell-size change observed in (Sr,La)RuO$_3$ is in sharp contrast to that of Sr(Ru,Rh)O$_3$, where a fairly small change occurs approximately $-0.1$\% in the volume -- a value comparable to the x-ray experimental error. 
This fact probably provides an advantage for the present investigation, because the influence of local-structure distortions is remarkably reduced over the whole Rh concentration, as the relation of the distortion to the magnetic order was frequently focused on as a significant issue in many experimental and theoretical studies \cite{PRB00TH,PRL99KY,JPSJ98TK,PRB97GC,JAP68JML,IC66AC,JPCM02HN,PRL01MSL}. 

\begin{figure}
\includegraphics[width=8cm]{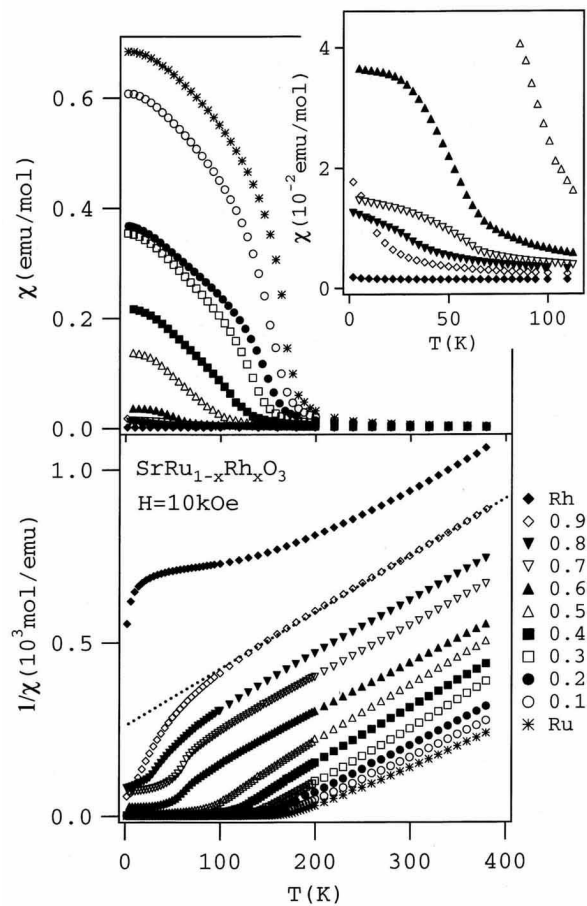}
\caption{Temperature dependence of the magnetic susceptibility of the polycrystalline SrRu$_{1-x}$Rh$_x$O$_3$ at 10 kOe on cooling ($\chi$ vs. $T$, upper panel), and the data in another form ($1/\chi$ vs. $T$, lower panel). The small panel shows an expansion of the low-temperature portion of the $\chi$ vs. $T$ plots. The dotted line indicates a representative for fit to the Curie-Weiss law.} 
\label{fig2} 
\end{figure}

\begin{figure}
\includegraphics[width=8cm]{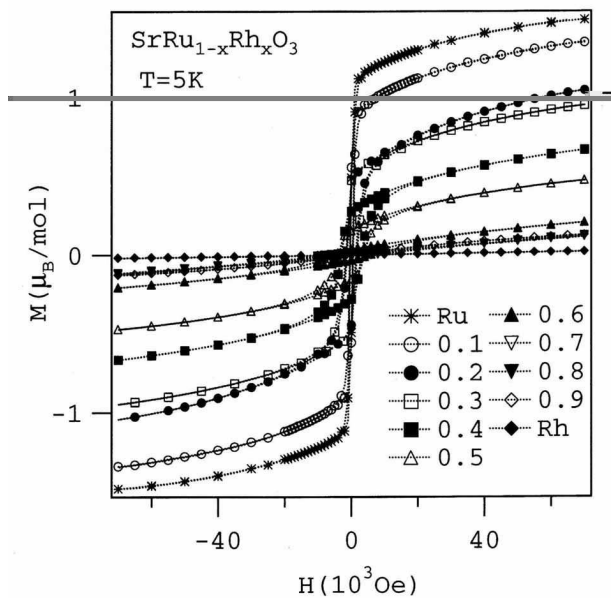}
\caption{Applied magnetic field dependence of the magnetization of the polycrystalline SrRu$_{1-x}$Rh$_x$O$_3$ at 5 K. Each symbol indicates every third points measured for each sample.}
\label{fig3}
\end{figure}

The magnetic data for Sr(Ru,Rh)O$_3$ are presented in Figs.\ref{fig2} and \ref{fig3}.
First, we shall focus on the end member SrRuO$_3$ ($x$=0.0). 
Shown in the figures, the magnetic susceptibility curve ($\chi$ vs. $T$) shows a clear ferromagnetic transition at $\sim$160 K and a spontaneous magnetic moment ($P_{\rm sp}$) of 0.82 $\mu_{\rm B}$ per Ru at 5 K, in good agreement with previous work \cite{JPSJ78AK,JPSJ76AK,JPSJ76AK2,JAP68JML,IC66AC,JPSJ99TK}, indicating the quality of the present sample. 
The $P_{\rm sp}$ was estimated by means of a least-squares fit to the magnetization data (Fig.\ref{fig3}):
A general empirical formula $M= P_{\rm sp} +aH^n$ was applied above approximately 10 kOe, where $a$ and $n$ were adjustable parameters \cite{PRB97TKN}. 

\begin{figure}
\includegraphics[width=8cm]{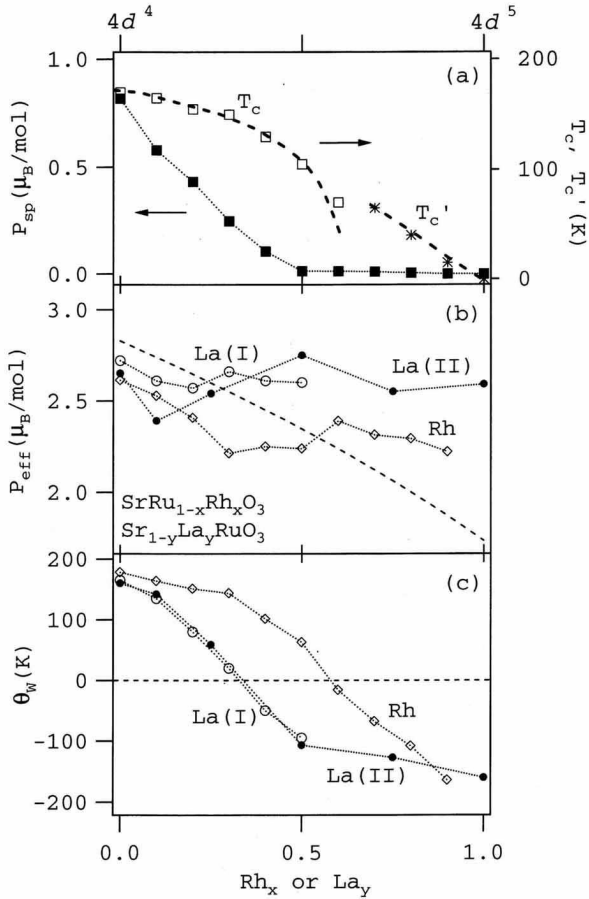}
\caption{Summary of magnetic parameters of SrRu$_{1-x}$Rh$_x$O$_3$ deduced from the experimental data, and comparison with those of the isoelectronic Sr$_{1-y}$La$_y$RuO$_3$ \cite{JPCM02HN,JSSC72RJB}; (a) Curie temperature $T_{\rm c}$, transition temperature of the small component $T_{\rm c}'$, and spontaneous magnetic moment $P_{\rm sp}$ measured at 5 K for SrRu$_{1-x}$Rh$_x$O$_3$. $P_{\rm sp}$ for La samples were not provided in the reports \cite{JPCM02HN,JSSC72RJB}. (b) Effective Bohr magneton $P_{\rm eff}$ per mole. The broken curve indicates an intermediate magnetic moment between $S=1/2$ and $S=1$, calculated by the formula $P_{\rm cal}=\sqrt{(1-x)P_{S=1}^2+xP_{S=1/2}^2}$, where $P_{S}=2\sqrt{S(S+1)}\mu_{\rm B}$. (c) Weiss temperature $\theta_{\rm W}$. }
\label{fig4}
\end{figure}

\begin{table*}
\caption{Magnetic parameters and the electronic-specific-heat coefficient of SrRu$_{1-x}$Rh$_x$O$_3$.}
\label{table1}
\begin{ruledtabular}
\begin{tabular}{llllll}
$x$ &$P_{\rm eff}$ ($\mu_{\rm B}$/mol) &$\theta_{\rm W}$ (K) &$P_{\rm sp}$ ($\mu_{\rm B}$/mol) &$\gamma$ (mJ$\cdot$mol$^{-1}\cdot$K$^{-2}$) &$T_{\rm c}$, $T_{\rm c}'$ (K)\footnotemark[1] \\
\hline
0	&2.613	&177.9 &0.818 &27.6	&170 \\
0.1	&2.527 &163.7 &0.579 &30.6	&165 \\
0.2	&2.406	&151.0 &0.432	&42.0	&155 \\
0.3	&2.212	&143.5 &0.247	&46.1	&150 \\
0.4	&2.248	&101.3 &0.105	&41.6	&130 \\
0.5	&2.237	& 62.94 &0.0126&32.6&105 \\
0.6	&2.388	&-15.92 &0.0123&15.7 &70 \\
0.7	&2.312	&-67.98 &0.0095&24.4 &65 \\
0.8	&2.291	&-108.5 &0.0043&45.6 &40 \\ 
0.9\footnotemark[2]	&2.219	&-164.2 &$<$0.001&	&15 \\
1\footnotemark[3]	&	&	  &$<$0.001&15.3	&0 \\
\end{tabular}
\end{ruledtabular}

\footnotetext[1]{$T_{\rm c}$ ($x\le 0.6$) and $T_{\rm c}'$ ($x>0.6$) are determined by a ``kink point'' method for the susceptibility data.}
\footnotetext[2]{the $\gamma$ is unable to be estimated.}
\footnotetext[3]{SrRhO$_3$ does not follow the Curie-Weiss law (Ref. 28). 
Previously, 7.6 mJ$\cdot$mol$^{-1}\cdot$K$^{-2}$ was reported for $\gamma$ of SrRhO$_3$ (Ref. 28). The disagreement possibly results from either an improvement of the sample quality (higher purity starting materials) or the choice of the temperature range to fit. The Debye temperature previously reported was miscalculated (Refs. 28 and 39). The correct calculations come within approximately a 70-K width, largely depending on the choice of the temperature range; 324--399 K and 316--362 K for $\Theta_{\rm D}$ of SrRhO$_3$ and Sr$_3$Rh$_2$O$_7$, respectively. }
\end{table*}

In order to quantitatively analyze the change in the magnetic susceptibility through the Rh substitution, the Curie-Weiss law was applied to the $1/\chi$ vs $T$ plots. 
The applied formula was $1/\chi(T)=3k_{\rm B}(T-\theta_{\rm W}) \cdot N_0^{-1} \cdot P^{-2}_{\rm eff}$, where $k_{\rm B}$, $N_0$, $P_{\rm eff}$, and $\theta_{\rm W}$ are the Boltzmann constant, Avogadro's constant, effective Bohr magneton, and Weiss temperature, respectively. 
As represented on the $x=$ 0.9 plot, a least-squares fit to the linear part (200 K $\le T \le$ 380 K) resulted in a fair evaluation of the Curie-Weiss parameters for all the samples except SrRhO$_3$, which is known not to follow the Curie-Weiss law \cite{PRB01KY}. 
The magnetic parameters of Sr(Ru,Rh)O$_3$ estimated in this study are listed in Table \ref{table1} and plotted in Fig.\ref{fig4} to make a comparison with those of the isoelectronic system (Sr,La)RuO$_3$. 

Shown in Fig.\ref{fig4}c, the Weiss temperature of Sr(Ru,Rh)O$_3$ switches sign at $x \sim 0.6$, accompanied by the disappearance of long-range ferromagnetic order.
This point ($x \sim$ 0.6) could be considered as a compositional critical point as was found for the related systems (Sr,Na,La)RuO$_3$ \cite{PRB00TH} and (Sr,Ca)RuO$_3$ \cite{PRL99KY,JPSJ98TK,PRB97GC}. 
A small component of ferromagnetic character persists beyond the critical point as seen in the susceptibility data for the $x=$ 0.6--0.9 samples (Fig.\ref{fig2}), and it is likely due to a formation of local magnetic clusters, where Ru is rich in the perovskite host. 
A correlation of the transition temperature ($T_{\rm c}'$) of the small ferromagnetic component with the Ru concentration supports this presumption (Fig.\ref{fig4}a). 
Alternatively, a small amount of magnetic impurities may be responsible for the ferromagnetic character. 
More will be said on this later. 

Shown in Fig.\ref{fig4}b, the effective Bohr magneton of Sr(Ru,Rh)O$_3$ undergoes a relatively small change between the ideal limits of $S=1$ for Ru$^{4+}$ and 1/2 for Rh$^{4+}$ over the whole range studied. 
However, for the samples above $x=0.6$, the values of $P_{\rm eff}$ are higher than the simple expectation indicated by the broken curve. 
A similar feature was also found in the La samples above $y=0.3$. 
This observation suggests that the $P_{\rm eff}$ enhancement is likely induced by the ferro to paramagnetic transition. 
In addition, an orbital contribution is possibly involved in the origin of the enhancement as discussed in the La substitution study \cite{JSSC72RJB}. 

As seen in Fig.\ref{fig4}a and Fig.\ref{fig4}c, the relatively quick suppression of the ferromagnetic order in (Sr,La)RuO$_3$ occurs more slowly in the less distorted Sr(Ru,Rh)O$_3$. 
For example, SrRu$_{0.6}$Rh$_{0.4}$O$_3$ shows ferromagnetic order ($T_{\rm c} \sim$ 100 K), while Sr$_{0.6}$La$_{0.4}$RuO$_3$ does not, although both are at the same doping level ($d^{4.4}$). 
Comparison between the isoelectronic and isostructural perovskites indicates that the GdFeO$_3$-type distortion substantially influences the ferromagnetic order as found in (Sr,Na,La)RuO$_3$ \cite{PRB00TH}. 

At the 4$d^5$ end, i.e.~SrRhO$_3$, a band-structure-calculation study predicts that a hypothetical cubic SrRhO$_3$ is likely to have a ferromagnetically spin-polarized ground state, in which the distortions are ideally free \cite{marai,PRB03DJS}.
This is in accord with a logical extension of the above consideration.
The prediction was tested on samples of (Sr,Ca)RhO$_3$ and (Sr,Ba)RhO$_3$, since the degree of distortions in these samples was systematically modified.
Unfortunately, the expected magnetic order in long range has not been found thus far \cite{PB03KY}. 

\begin{figure}
\includegraphics[width=8cm]{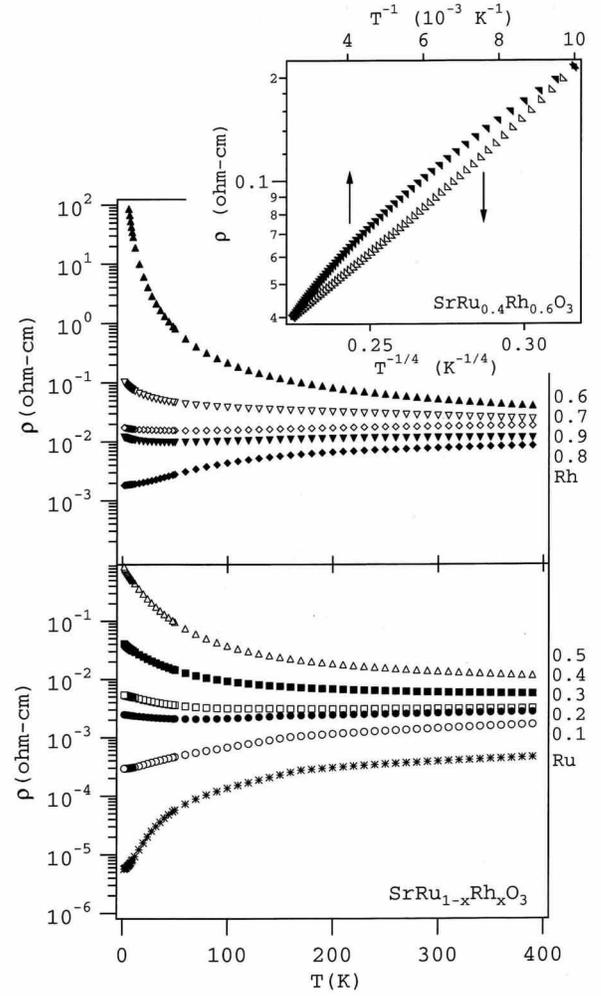}
\caption{Temperature dependence of the electrical resistivity of the polycrystalline SrRu$_{1-x}$Rh$_x$O$_3$. 
The data for $x=$ 0.0--0.5 samples are shown in the lower panel, and the others are in the upper panel.
Top panel: comparison between two types of plots of the electrical resistivity data at $x=0.6$.}
\label{fig5}
\end{figure}

\begin{figure} 
\includegraphics[width=8cm]{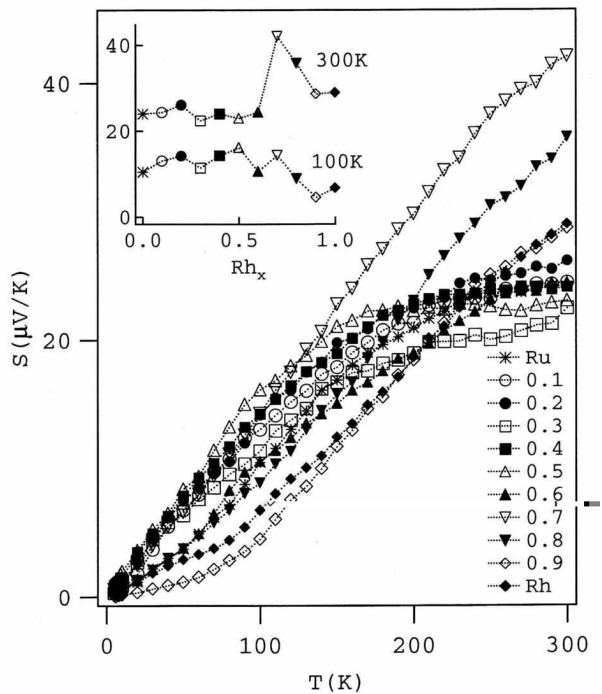}
\caption{Thermoelectric power of the polycrystalline SrRu$_{1-x}$Rh$_x$O$_3$. Two sets of data at 100 K and 300 K are compared in the inset.}
\label{fig6}
\end{figure}

\begin{figure} 
\includegraphics[width=8cm]{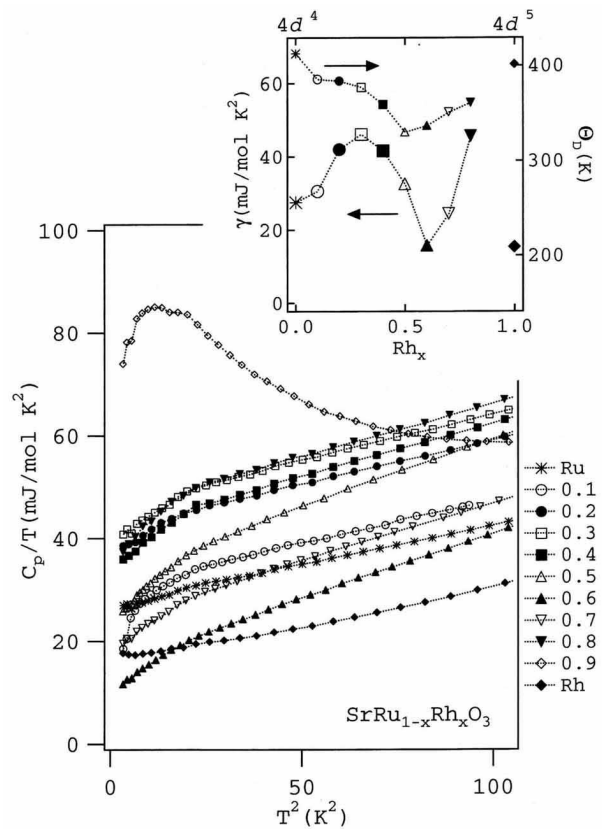}
\caption{Specific heat of the polycrystalline SrRu$_{1-x}$Rh$_x$O$_3$. 
The data are plotted in the $C_{\rm p}/T$ vs. $T^2$ form. 
Dotted curves are guides to the eye. 
Fit to the linear parts by the formula (Eq.\ref{eq1}) yields electronic specific-heat coefficient ($\gamma$) and Debye temperature ($\Theta_{\rm D}$) in preliminary sense, those are plotted versus the Rh concentration in the small panel.}
\label{fig7}
\end{figure}

\begin{figure} 
\includegraphics[width=8cm]{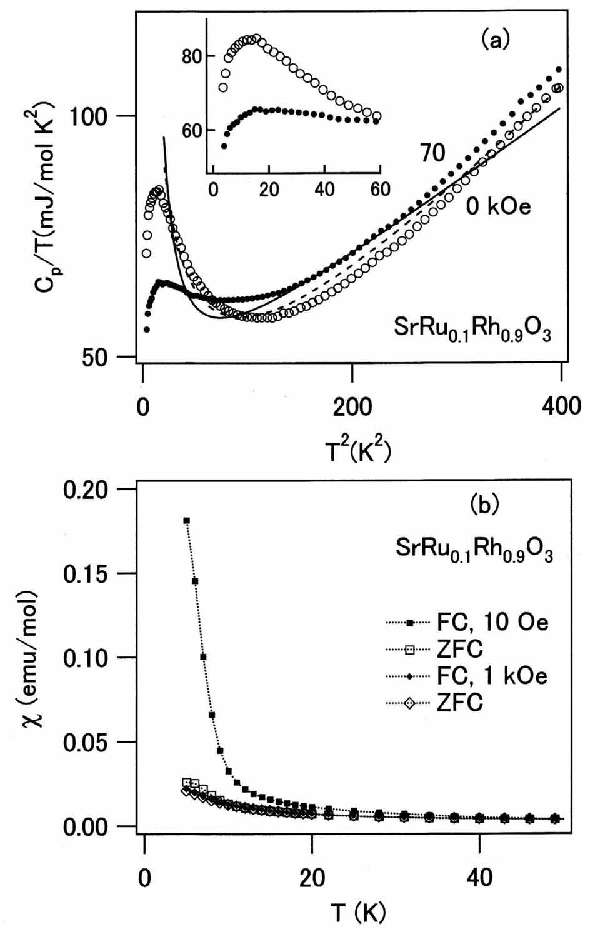}
\caption{(a) Specific heat of SrRu$_{0.1}$Rh$_{0.9}$O$_3$ measured in the magnetic field 70 kOe (solid circles). The inset shows an expansion of the low-temperature portion. Attempts to fit the probable magnetic contribution were made with the zero-field data (open circles). The data above the peak temperature is unlikely to follow either an empirical form (dotted curve) or a form including a Schottky term (solid curve). Details are in the text. (b) Temperature dependence of the magnetic susceptibility at the low fields, 10 Oe and 1 kOe, on heating to 300 K after cooling the sample without an applied magnetic field (zero field cooling, ZFC) and then on cooling in the field (field cooling, FC). }
\label{fig8}
\end{figure}

Electronic transport properties were measured on the polycrystalline Sr(Ru,Rh)O$_3$ samples; electrical resistivity, thermopower, and specific-heat data are shown in Figs.\ref{fig5}--\ref{fig7}, respectively. 
The data for Sr(Ru,Rh)O$_3$ indicate that the resistivity has a large composition dependence, as more than 6 orders of magnitude change are observed from the metallic to semiconducting state (as shown in Fig.\ref{fig5}). 
The end member, SrRuO$_3$, has a fairly metallic character with a kink at $\sim$ 160 K, in good agreement with previous reports \cite{PRL02LC,PRB00TH,PRB98APM,JPCM96LK}. 
The metallic character gradually transforms to semiconductor-like with increasing Rh concentration up to $x= 0.6$, and then the resistive character reverts back to being metallic for further Rh substitution. 
Although the whole resistivity data are probably somewhat complicated by the polycrystalline nature of samples, including resistive grain boundaries, the remarkably large dependence on composition reflects mainly a disorder effect related to the distribution of Rh and Ru at the perovskite B-site. 
This is strongly suggested by the resistive character becoming larger at approximately the center of the composition range. 
In order to investigate the probable disorder effect further, the resistivity data at $x=$ 0.6 were plotted into two independent forms as shown in the top panel in Fig.\ref{fig5}. 
As expected, the data follow closely to a variable-range-hopping form (logarithmic resistivity $\sim  T^{-1/4}$), much more than the alternative form ($\sim  T^{-1}$), which suggests the disorder effect is predominant in the semiconducting state. 

Although the electrical character of Sr(Ru,Rh)O$_3$ varies tremendously, the thermoelectric property is rather featureless (Fig.\ref{fig6}). 
All the Seebeck data are consistent with a low-carrier density metallic system, and the positive Seebeck coefficient indicates that the majority carrier is holelike. 
It is known that the thermopower is more sensitive than resistivity to changes in the electronic structure of materials.  
Why the resistivity (varying by 6 orders of magnitude) is so sensitive to chemical disorder and the thermopower is not is unclear.  
However, for the samples above the critical concentration ($x \sim 0.6$), the thermopower remains relatively linear in temperature above 100 K, in good accord with free-electron diffusion thermopower of metals.  
For samples below $x = 0.6$, the thermopower tends to saturate at high temperatures. 
Nonlinearity in the diffusion thermopower, as like those, is often observed in materials containing transition metals \cite{Taylor72RDB}. 

The specific-heat data (Fig.\ref{fig7}) were attempted to be analyzed by two independent parameters $\gamma$ (electronic-specific-heat coefficient) and $\Theta_{\rm D}$ (Debye temperature) in the low-temperature limit ($T \ll \Theta_{\rm D}$). 
The analytic formula employed here was  
\begin{eqnarray}
{C_{\text v} \over T} = \gamma + {12\pi^4 \over 5}rN_0k_{\text B}\Big({1 \over \Theta_{\text D}^3}\Big)T^2,
\label{eq1}
\end{eqnarray}
where $r$ was the number of atoms per formula unit. 
Since a magnetic contribution is expected to appear in addition to the two terms in Eq.\ref{eq1}, this attempt is therefore preliminary \cite{Springer85TM,AP00TM}. 
However, we expected the attempt for a valuable achievement as was done in many specific-heat studies on magnetic materials. 
The difference between $C_{\rm p}$ and $C_{\rm v}$ was assumed insignificant in the temperature range studied.
The values of $\gamma$ and $\Theta_{\rm D}$ for the samples of Sr(Ru,Rh)O$_3$ were then obtained by a least-squares method with the linear part ($C_{\rm p}/T$ vs. $T^2$) between 30 K$^2$ and 100 K$^2$ (5.4 K$<T<$10 K). 
The parameters are plotted versus Rh concentration in the small panel of Fig.\ref{fig7}. 
It should be noted that the points for the $x=0.9$ sample are missing, because its surprising nonlinear specific heat did not allow them to be extracted from the data in the usual way (the up turn was reproducibly observed). 

From the fit, we find $\gamma = 28$ mJ$\cdot$mol$^{-1}\cdot$K$^{-2}$ for SrRuO$_3$, which is in good agreement with the established data range from 29--36 mJ$\cdot$mol$^{-1}\cdot$K$^{-2}$, while the Debye temperature ($\Theta_{\rm D}=412$ K) is slightly higher than the reported values by approximately 20 to 40 K \cite{PRB96PBA,PRB99JO,JAP99CSA,JPSJ98TK,JAP97MS,PRB97GC}. 
This small discrepancy may result from trace chemical impurities contained in the present sample of SrRuO$_3$ or the sample is slightly off stoichiometry, which can occur at the extreme heating conditions. 
Alternatively, a possible magnetic contribution might influence the fit to the data somewhat because there are no magnetic terms in Eq.\ref{eq1}, if the high-pressure heating varied slightly the degree of local-structure distortions which are sensitive to the magnetism \cite{PRL01MSL}. 

The $\gamma$ curve in the small panel of Fig.\ref{fig7} shows a rather complicated feature, while the Debye-temperature curve shows a simple V shape reflecting the lattice change. 
The $\gamma$ value reaches a minimum at $x \sim 0.6$, where the long-range magnetic order disappears, and becomes increasingly prominent at $x\sim 0.9$.
This is also the concentration where the Curie-Weiss-type paramagnetism ($1/\chi \sim T$) transforms to the unconventional type ($1/\chi \sim T^2$) \cite{PRB01KY}. 
These results suggest that $\gamma$ is likely being coupled to the magnetic properties of the series. 

Through the present investigation of Sr(Ru,Rh)O$_3$, we measured a variety of properties of the polycrystalline samples, and found the specific-heat data were perhaps the most intriguing. 
The specific-heat data for SrRu$_{1-x}$Rh$_x$O$_3$ clearly imply that a magnetic factor is substantial in the origin of the anomalous $\gamma$ feature. 
We then investigated the $\gamma$ feature of the closely related system (Sr,Ca)RuO$_3$ to make a comparison with the present data: We found that the phenomenology of the $\gamma$ curve for (Sr,Ca)RuO$_3$ does not match at all with that for Sr(Ru,Rh)O$_3$ \cite{SrRhO3gamma}. 
An enhancement of $\gamma$ in the vicinity of the compositional critical point was more commonly observed for the ferromagnetic systems, including Y(Co,Al)$_2$ \cite{JPSJ90HW}, Lu(Co,Al)$_2$ \cite{PSSB90HW}, and NiCu alloys \cite{JPF83JWL}, rather than the decrease of $\gamma$. 
A possible picture invoking the variation of the density of states at the Fermi level is unlikely to account for the $\gamma$ suppression in Sr(Ru,Rh)O$_3$, because the degree of structural variation is rather small (Fig.\ref{fig1}). 
A band-filling picture does not seem adequate, either \cite{marai}. 
The most likely picture for the $\gamma$ decrease is due to a randomness effect. 
As suggested by the electrical resistivity study, the B-site randomness plays a significant role in the electronic system of the solid-solution Sr(Ru,Rh)O$_3$. 
The $\gamma$ in the system decreases probably due to an influence of the disorder as well as decreasing of the electrical conductivity. 
It is, however, still noteworthy that the $\gamma$ remains large (more than 15 mJ$\cdot$mol$^{-1}\cdot$K$^{-2}$), although the electrical resistivity is rather large (more than 100 ohm$\cdot$cm). 
Currently, we have not yet reached a clear conclusion on how $\gamma$ gets suppressed at the ferromagnetic critical point, however, the data suggest that the randomness contribution is likely essential in the origin.
Further studies focused on the possible association between the randomness and the magnetic order would be of interest. 

The anomaly in the specific heat of the $x = 0.9$ sample is also probably magnetic in origin, and in order to investigate this feature, three possible forms were tested by a least-squares method; (1) The Schottky term $\varepsilon/T^3$, where $\varepsilon$ was an adjustable factor, was added linearly to Eq.\ref{eq1} \cite{JPF83JWL,JPSJ90HW}. 
(2) Empirical magnetic terms $\delta/T$ and (3) $\alpha T^2\log T$, where $\delta$ and $\alpha$ were adjustable factors, were attempted as well \cite{JPF83JWL,JPSJ90HW}. 
Accordingly, none of the three attempts produced a good fit to the data; The curves providing the best fits are shown in Fig.\ref{fig8}a.  
The curve for (3) was not shown because the parameters were unphysical. 
The fact that the regular forms do not satisfy the $x = 0.9$ data (Fig.\ref{fig8}a) suggests unconventional magnetism \cite{PRB03DJS}. 
A corresponding $\gamma$ peak was not observed in (Sr,Ca)RuO$_3$\cite{JPSJ98TK,PRB97GC}, so the origin of the anomaly in the $x=$ 0.9 data is, therefore, of interest. 

The specific heat of the $x = 0.9$ sample was reinvestigated in magnetic field. 
The measurement was conducted at 70 kOe, and the data were plotted with the zero-field data in Fig.\ref{fig8}a. 
As shown, the peak is clearly suppressed and the linear part slightly shifts in parallel. 
These facts clearly indicate the manifest peak at $x = 0.9$ results from a magnetic origin, and the linear part includes magnetic contributions in addition to the regular electronic term. 
The analytic formula in Eq.\ref{eq1} is therefore insufficient to clarify the specific heat data of SrRu$_{1-x}$Rh$_x$O$_3$; magnetic terms should be appended to Eq.\ref{eq1} for a more rigorous analysis.
Furthermore, an expected peak shift in the magnetic field, if spin-glass or magnetic-cluster glass contribution is significant, is not clearly seen (inset in Fig.\ref{fig8}a) \cite{PRB02AL}. 
The field-induced suppression suggests a spin-fluctuations contribution, however an attempt to fit to the $C_{\rm p}$ data by the analytical formula for spin-fluctuations was not at all successful; it was worse than the previous case (the curve to fit is not shown) \cite{PRB01KY}. 

In order to investigate the $x = 0.9$ sample further, low-field-magnetic susceptibility was measured; the data are presented in Fig.\ref{fig8}b. 
Thermomagnetic hysteresis was clearly observed at 10 Oe below about 20 K, suggesting a magnetically glassy character. 
The glassy feature may be due to freezing of ferromagnetic clusters, as suggested by the high-field susceptibility data (Fig.\ref{fig2}), or a possible spin-glass transition. 
The data above, unfortunately, did not provide a sufficient information to identify the origin of the peculiarities of $C_{\rm p}$ at $x = 0.9$. 
Further measurements will be needed to clarify the nature of the characteristic peak. 

In summary, we have investigated the electronic transport and magnetic properties of the full-range-solid solution SrRu$_{1-x}$Rh$_x$O$_3$ ($0\le x \le1$), which is a variable-electron configuration system from 4$d^4$ to 4$d^5$ on the perovskite-structure basis. 
We found that the $\gamma$ behavior for SrRu$_{1-x}$Rh$_x$O$_3$ does not match even qualitatively with that for the intimately related (Sr,Ca)RuO$_3$. 
The substitution randomness probably contributes significantly to the $\gamma$ decrease around the compositional critical point ($x \sim 0.6$),
which unavoidably prevents us from studying a true magnetic contribution to the $\gamma$ feature. 
Meanwhile, the data suggest that magnetic contributions play a central role in the specific heat in the vicinity of $x=0.9$. 
Possible formation of Ru-rich magnetic clusters in the perovskite host may be responsible for the observations, however the features of $C_{\rm p}$ were not fully explained by the model alone. 
The unconventional magnetism of SrRhO$_3$ is also possibly associated with the peculiarities of $C_{\rm p}$ at $x = 0.9$. 
Further experimental studies, as well as theoretical considerations, would be significant in expanding our insight into the ferromagnetism of Sr(Ru,Rh)O$_3$. 

\acknowledgments
We wish to thank M. Akaishi and S. Yamaoka for their advice on the high-pressure experiments, and M. Arai for productive discussion. 
This research was supported in part by the Superconducting Materials Research Project, administrated by the Ministry of Education, Culture, Sports, Science and Technology of Japan.  K.Y. was supported by the Domestic Research Fellowship, administrated by the Japan Society for the Promotion of Science.

\pagebreak


\begin{references}
\bibitem{SrO2}
SrO$_2$ was prepared as follows. 
SrCl$_2$-6H$_2$O was dissolved in $\sim$300 ml of water, followed by dropwise of ammonia solution and H$_2$O$_2$ solution.
The precipitation was washed with water after filtered, and then dried in oxygen for 3 hours, resulting in white powder. 
Thermogravimetric and x-ray diffraction analysis characterized the powder to be SrO$_2$.
\bibitem{SrRhO3gamma}
The two reports (Refs. 36 and 37) for $\gamma$ of (Sr,Ca)RuO$_3$ are entirely consistent except one ($x=0.25$) of five points in Cao's data.
\end{references}
\end{document}